\documentclass[graybox]{svmult}

\usepackage{aas_macros}
\usepackage{helvet}         
\usepackage{courier}        
\usepackage{type1cm}        
\usepackage{makeidx}         
\usepackage{graphicx}        
\usepackage{multicol}        
\usepackage[bottom]{footmisc}

\makeindex             

\begin{document}

\title*{Signatures of magnetic reconnection in solar eruptive flares: A multi-wavelength perspective}
\titlerunning{A multi-wavelength perspective of solar eruptive flares} 

\author{Bhuwan Joshi, Astrid Veronig, P. K. Manoharan, and B. V. Somov}
\institute{Bhuwan Joshi \at Udaipur Solar Observatory, Physical Research Laboratory, Udaipur 313 004, India \email{bhuwan@prl.res.in}
\and  Astrid Veronig \at IGAM/Institute of Physics, University of Graz, Universit$\ddot{a}$tsplatz 5, A-8010 Graz, Austria \email{astrid.veronig@uni-graz.at}
\and P. K. Manoharan \at Radio Astronomy Centre, Tata Institute of Fundamental Research, Udhagamandalam (Ooty) 643 001, India \email{mano@ncra.tifr.res.in}
\and Boris V. Somov \at Astronomical Institute, Moscow State University, Universitetskij Prospekt 13, Moscow 119992 \email{somov@sai.msu.ru}.}

\authorrunning{Joshi et al.}
%
%
\maketitle

\abstract{
In this article, we review some key aspects of a multi-wavelength flare which have essentially contributed to form a standard flare model based on the magnetic reconnection. The emphasis is given on the recent observations taken by the Reuven Ramaty High Energy Solar Spectroscopic Imager (RHESSI) on the X-ray emission originating from different regions of the coronal loops. 
We also briefly summarize those observations which do not seem to accommodate within the canonical flare picture and discuss the challenges for future investigations\footnote
{{\bf \it Published in} {\sc Multi-scale Dynamical Processes in Space and Astrophysical Plasmas}, Astrophysics and Space Science Proceedings, Springer-Verlag Berlin Heidelberg, {\bf 33}, 29–-41 (2012).}.}


\section{Introduction}
\label{sec:1}

\begin{figure}[t]
\centering
\includegraphics[scale=0.9]{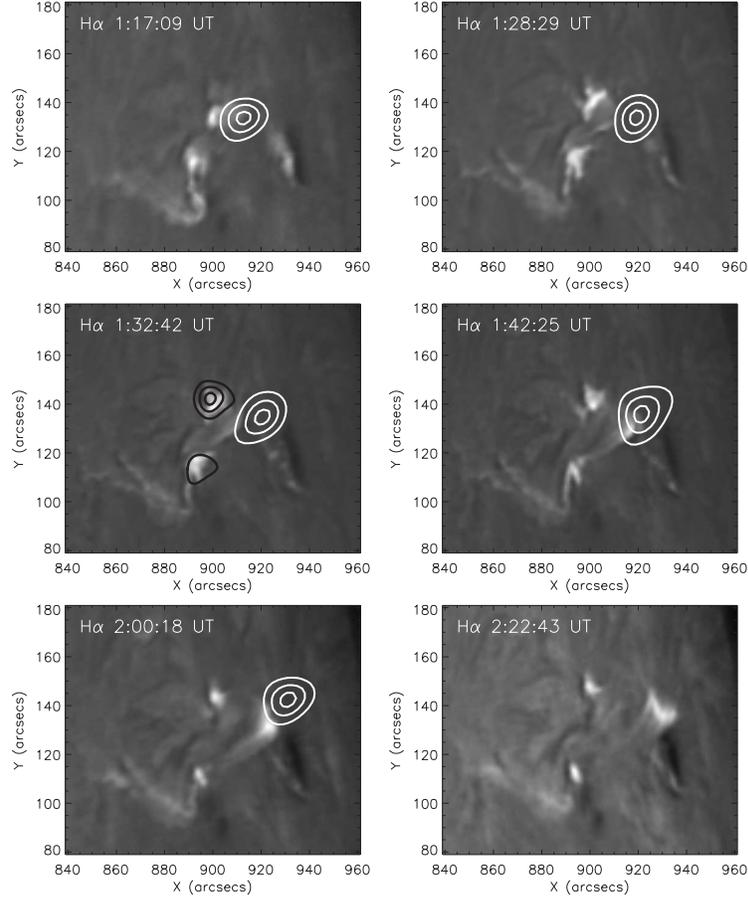}
\caption{Time evolution of a large eruptive flare of 2B/X2.7 class that occurred in active region NOAA 10488 on November 3, 2003 at a location N10W83, i.e., close to the west limb of the Sun. The H$\alpha$ filtergrams, observed from ARIES Solar Tower Telescope, are overlayed by RHESSI X-ray images at 10-15 keV (white contours) and 50-100 keV (black contours; only in the third panel). Figure adopted from \cite{joshi2007}.}
\label{fig:1}       
\end{figure}

Solar flares are the most striking explosive form of solar activity. A flare is characterized by a sudden catastrophic release of energy in the solar atmosphere. In tens of minutes energy in excess of $10^{32}$ erg is released. Mostly flares occur in solar active regions; being more frequent at the locations where the active region is rapidly evolving \cite{hagyard1984}. The frequency and intensity of solar flare occurrence follow the 11-year sunspot cycle \cite{joshi2005,joshi2006}.

Flares manifest their signatures in a wide range of electromagnetic spectrum, from radio to $\gamma$-rays, and involve substantial mass motions and particle acceleration. Emission in these wavelengths originates from the atmospheric layers of the Sun extending from the chromosphere to the corona. In general, flares are not visible in the photosphere except in some exceptionally high energy and impulsive events known as white light flares. Now it is well known that the energy released during flares is stored in the corona prior to the event in the form of stressed or non-potential magnetic fields. Magnetic reconnection has been recognized as the fundamental process responsible for the changes in the topology of magnetic fields, as well as the rapid conversion of stored magnetic energy into heat and kinetic energy of plasma and particles during a flare.


In section \ref{sec:2}, we provide an observational overview of solar flares and discuss some key aspects of flare development which form the basis for the standard flare model. With high resolution observations at temporal and spectral domains, now we have a clearer view on these multi-wavelength flare components. Recent observations, mainly inspired by Reuven Ramaty High Energy Solar Spectroscopic Imager (RHESSI) \cite{lin2002}, have further revealed many new multi-wavelength aspects of flare emission which have changed our ``standard" understanding on flares. These new results are summarized in section \ref{sec:3}. 

\section{Overview of multi-wavelength phenomena}
\label{sec:2}

\subsection{Confined and eruptive flares}
\label{subsec:2a}

The very early observations of solar flares in soft X-ray (SXR) wavelengths from the {\it Skylab} mission in 1973--1974 established two morphologically distinct classes of flares: confined and eruptive events \cite{pallavicini1977}. The confined flares show brightening in compact loop structures with little large-scale motion.  They are generally modeled  in terms of energy release within a single static magnetic loop and are thus referred to as single-loop, compact or point flares.  
The second category comprises the long duration events (LDE) which are 
eruptive in nature. They are 
accompanied by an arcade of loops and show more strong association with coronal mass ejections (CMEs). It is worth to mention that the temporal evolution of CME and flare signatures in eruptive events suggests that both phenomena have a strongly coupled relationship but not a cause--effect one \cite{zhang2001}.
The H$\alpha$ observations reveal that eruptive events are almost always associated with chromospheric brightenings in the form of long, bright parallel ribbons. Therefore, LDE flares are also referred to as two-ribbon flares. This two-element classification of solar flares is also broadly reflected in the X-ray observations of stellar flares \cite{pallavicini1990,pandey2012}.

\subsection{Time evolution of an eruptive flare}

\label{subsec:2b}
A solar flare is a multi-wavelength phenomenon. Therefore in order to have a complete understanding of its temporal evolution we need to look at the time profiles observed at different wavelengths. 
However, it has been observed that there could be subtle activities at the flare location before its onset. In the following, we discuss different aspects of flare evolution.

\subsubsection{Pre-flare activity and precursor phase}
The pre-flare activity refers to the very earliest stage of flare which is elusive to recognize even in SXRs \cite{joshi2011}. The activities in this initiation phase can be seen at longer wavelength such as H$\alpha$, ultraviolet (UV), and extreme ultraviolet (EUV). The observations of subtle changes in the configuration of EUV loops and localized brightenings at this early phase can provide important clues about the triggering mechanism of the eruption \cite{chifor2007,joshi2011}. 
It has been suggested that the pre-flare brightening may occur as a result of slow magnetic reconnection and provide a trigger for the subsequent large-scale eruption \cite{moore1992,chifor2007}.

Many flares show slow and gradual enhancement in SXRs before the onset of the impulsive energy release, referred to as the X-ray precursor phase \cite{tappin1991}. This early phase mainly corresponds to small-scale brightening in UV to SXR wavelengths \cite{chifor2007,kim2008,joshi2011}. The precursor flare brightening mostly occurs in the neighborhood of the main flare location \cite{farnik1996}. However, usually precursor and main flare locations do not exactly coincide \cite{farnik1998,warren2001}. Some studies recognize the precursor phase brightenings as the evidence for distinct, localized instances of energy release which play a significant role in destabilizing the magnetic configuration of active region leading to eruption and large-scale magnetic reorganization \cite{chifor2007,joshi2011}.

\subsubsection{Impulsive phase}

\label{subsec:2b-3}
The primary energy release takes place during the impulsive phase which lasts from tens of seconds to tens of minutes. This phase is marked by emission in hard X-rays (HXR), non-thermal microwaves and in some cases also $\gamma$-rays and white-light continuum, showing evidence of strong acceleration of both electrons and ions. These radiations are further supplemented by strong enhancement of emissions in chromospheric lines (e.g., H$\alpha$), ultraviolet and extreme ultraviolet. The impulsive phase is mainly characterized by the flare signatures at chromospheric layers where the feet of magnetic loops are rooted at both sides of the polarity inversion line.  Morphologically, flare brightenings at this region are termed as ``footpoints" or ``ribbons" detected in HXRs and H$\alpha$ observations, respectively. With the upward expansion of the arcade of loops, the two parallel flare ribbons (or HXR footpoints) separate from each other during the impulsive phase and later. The HXR emission from the footpoints of flaring loops is traditionally viewed in terms of the thick-target bremsstrahlung process in which the X-ray production at the footpoints of the loop system takes place when high-energy electrons, accelerated in the coronal reconnection region, come along the guiding magnetic field lines and penetrate the denser transition region and chromospheric layers \cite{brown1971,Syrovatskii1972}. During very high energetic events, the photospheric Doppler enhancements have been reported which are co-spatial with H$\alpha$ flare ribbons \cite{kumar2006,venkat2008}. 

In the impulsive phase, the time evolution of the spectral index of the non-thermal part of the photon spectrum and non-thermal flux show an interesting pattern known as {\it soft-hard-soft} spectral evolution. The HXR spectra of flares often initially show a steep spectral slope (soft), which flattens at the peak of the flare (hard), and then becomes steeper again (soft) in the decay phase of the flare \cite{benz1977}. With RHESSI data, the soft-hard-soft spectral evolution has been determined with much better accuracy. RHESSI observations clearly recognize that the spectral {\it soft-hard-soft} behavior in rise-peak-decay phase is followed not only in overall flare development, but even more pronounced in sub-peaks \cite{grigis2004}. This anti-correlation between spectral index and flux is generally interpreted as a signature of the acceleration process with each non-thermal peak representing a distinct acceleration event of the electrons in the flare \cite{grigis2004,joshi2011}.
Although the {\it soft-hard-soft} spectral evolution is very common, it does not apply to all flares. Some flares also exhibit {\it soft-hard-harder} patterns in which the spectrum continues to become harder throughout the flare evolution \cite{cliver1986,kiplinger1995}. This pattern is more commonly seen in microwave spectra than in HXR observations \cite{silva2000}. This is mainly reported in gradual HXR events, in particular those which are associated with solar proton events \cite{cliver1986,kiplinger1995,saldanha2008}. The {\it soft-hard-harder} pattern is attributed to extended phases of acceleration in large flares. Some events of this class also exhibit the long-lived high energy coronal sources \cite{krucker2008}.

\subsubsection{Gradual phase}
\label{subsec:2b-4}

The gradual phase of a solar flare is best described by SXR time profile. During the impulsive onset of HXR emission (see section \ref{subsec:2b-3}), the SXR gradually builds up in strength and peaks a few minutes {\it after} the impulsive emission. This implies that the long-lived and gradual SXR emission is a delayed effect of the impulsive onset of HXR radiation. This phase is characterized by the formation of loops (and arcade of loops in large flares) which emit in SXRs and EUV, indicating the presence of hot plasma ($\sim$10--20 MK) inside them. The process of filling of hot plasma in coronal loops is termed as {\it chromospheric evaporation} \cite{neupert1968,lin1976,milligan2006a,milligan2006b,veronig2010,nitta2011}. The chromospheric plasma is rapidly heated and compelled to spread out in the coronal loops primarily by the energy deposition of energetic electrons accelerated at the magnetic reconnection site in the corona \cite{lin1976,veronig2010}. Thermal conduction from the corona may also play a role in heating the chromospheric plasma \cite{antiochos1978,zarro1988,battaglia2009}. The flare loop system exhibits a gradient in temperature with outermost loops being the hottest \cite{forbes1996}. 

The {\it Yohkoh/ SXT } observations detected cusp-shaped structure above the hottest outer loops in many LDE flares \cite{tsuneta1992,tsuneta1996,forbes1996}. The soft X-ray arcades along with the cusp resemble the general geometry of large-scale magnetic reconnection \cite{tsuneta1996}. In the later stages, as the loops begin to cool, the arcade becomes visible in lower temperature emissions such as EUV and H$\alpha$ \cite{schmieder1995,forbes1996,uddin2003,vrsnak2006}. 
Both conduction and radiation may contribute to the cooling process which essentially depends on flare loop length and plasma parameters \cite{culhane1970,aschwanden2001}. Another important observational feature of gradual phase is the continual downflow of lower temperature plasma observed in H$\alpha$ visible along the leg of arcade. This H$\alpha$ downflow (also termed as `coronal rain') is the result of draining of cool plasma due to gravity \cite{brosius2003}.


\subsection{``Sigmoid-to-arcade" development}
\label{subsec:2c}
Sigmoids are S-shaped (or inverse S-shaped) coronal features, mainly identified in SXR images, in the form of a region of enhanced emission \cite{rust1996,manoharan1996,canfield2007}. In a few studies, sigmoid structures have been reported in EUV \cite{liuc2007,liu2010} and in one of the events even in HXRs \cite{ji2008}. They are often composed of two opposite J-like bundles of loops which collectively form an S-shape feature \cite{canfield2007}. Sigmoid regions are considerably more likely to be eruptive than non-sigmoidal sites \cite{hudson1998,glover2000}. With the onset of an eruptive flare, the region is enveloped by arcades or cusped loops (see section \ref{subsec:2b-4}). Thus the ``sigmoid to arcade" development is suggestive of large-scale magnetic reconnection driven by the eruption \cite{sterling2000,moore2001}.

\subsection{Standard flare model}
\label{subsec:2d}
The standard flare model, also known as CSHKP model, recognizes that the evolution of flare loops and ribbon can be understood as a consequence of the relaxation of magnetic field lines stretched by the ejection of plasma \cite{carmichael1964,sturrock1966,hirayama1974,kopp1976}. The magnetic reconnection has been identified as the key process which releases sufficient magnetic energy on short time scales to account for the radiative and kinetic energies observed during an eruptive event \cite{priest2002}. In this picture, the rise of the loop system as well as the footpoint (or ribbon) separation reflect the upward movement of the magnetic reconnection site during which field lines, rooted successively apart from the magnetic inversion line, reconnect. This picture successfully explains the apparent motions of flare loops and ribbons along with the multi-wavelength view of the loop system with the hottest one located at the outermost region. 

The discovery of the HXR source located {\it above} the soft X-ray flare loops (known as ``above-the-looptop" source) by {\it Yohkoh} was an important landmark in the history of solar flare observations \cite{masuda1994}. This new kind of HXR emission raised great interest as it is believed to occur closest to the particle acceleration region associated with the magnetic reconnection site. Essentially the {\it Yohkoh} discoveries of the HXR above-the-looptop source along with the SXR cusp (see section \ref{subsec:2b-4})  confirmed the role of magnetic reconnection in the standard flare model. Here it is worth mentioning that the above-the-looptop source is still a rarely observed feature \cite{ishikawa2011}. However, due to high sensitivity and broad energy coverage of RHESSI, the HXR emission from the looptop has now become a well known phenomenon and coronal HXR sources are detected in all phases of solar flares \cite{krucker2008a,krucker2008b}. In a few events, coronal HXR sources are even observed before flare impulsive phase \cite{lin2003,veronig2006}. RHESSI observations have also identified a new class of coronal sources that show strong looptop HXR emission without significant footpoint emissions \cite{veronig2004,veronig2005,joshi2011}. 

\section{Beyond the ``standard" observations}
\label{sec:3}

\begin{figure}[t]
\centering
\includegraphics[scale=0.80]{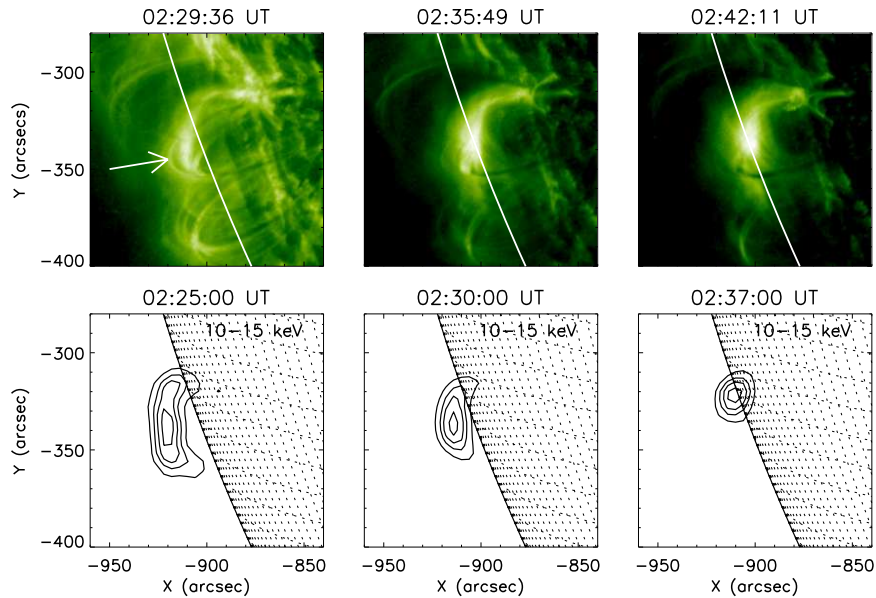}
\caption{Large scale contraction of coronal loops observed for $\sim$11 minutes in an M7.6 flare that occurred in active region NOAA 10486 on October 24, 2003 at a location S19E72 \cite{joshi2009}. The contraction of coronal loops can be readily seen in TRACE EUV images at 195~\AA~(top panels) and RHESSI X-ray images at 10-15 keV (bottom panels).}
\label{fig:2}       
\end{figure}

\subsection{Contraction of coronal loops}
\label{subsec:3a}
RHESSI observations have discovered that during the early impulsive phase of the flare, coronal loop system undergoes an altitude decrease or contraction before showing the ``standard" behavior of apparent outward expansion. This phenomenon of loop contraction or shrinkage, first reported in an M1.2 class flare on April 15, 2002 \cite{sui2003}, has now been confirmed in several events of different classes \cite{sui2004,veronig2006,joshi2007,reznikova2010}. Motivated by these RHESSI findings, the loop contraction was examined in other wavelength images and similar kind of descending loop motion was detected in EUV \cite{li2006,joshi2009,liu2009} and microwave images \cite{li2005}. 
The study of an M7.6 class flare on October 24, 2003 is perhaps the best example of the loop contraction which was observed for $\sim$11 minutes and identified in HXRs (up to $\sim$25-50 keV) and EUV images \cite{joshi2009}. Some observations also apparently imply that the contraction motion of flaring loops may be the result of the relaxation of the sheared magnetic field \cite{ji2007}.

\subsection{Converging motion of footpoints}
\label{subsec:3b}
In some events, the converging motion of footpoint sources (i.e., decrease in footpoint separation) has been observed at the early stage which temporally matches with the phase of contraction of coronal loops \cite{ji2006,ji2007,joshi2009,joshi2011}. 
A detailed analysis of footpoint motions was carried out for X10 class flare on October 29, 2003 \cite{liu2008}. In this study it was found that the two conjugate footpoints first move toward and then away from each other, mainly parallel and perpendicular to the magnetic inversion line, respectively. Further the transition of these two phases of footpoint motions coincides with the direction reversal of the motion of the looptop source. An interpretation of this new kind of motions of looptop and footpoint sources is proposed in terms of {\it rainbow reconnection model} \cite{somov1986} which essentially builds on the idea of three-dimensional magnetic reconnection at a magnetic separator in the corona during the phase of {\it shear relaxation} of coronal loops \cite{joshi2009,somov2010}.

\subsection{Double coronal sources}
\label{subsec:3c}
The RHESSI observations of three homologous flares that occurred between April 14-16, 2002 revealed the appearance of a coronal X-ray source besides the X-ray looptop emission \cite{sui2003,sui2004} which was formed at higher altitudes. 
Out of these, one event clearly shows a cusp-shaped flare loop in the rise phase \cite{sui2003,sui2005}. When the impulsive rise in HXRs ($>$25 keV) began, the cusp part of the coronal source separated from the underlying flare loop, forming two HXR emitting sources. The two sources exhibit energy dependent structures with the emissions at higher energies coming from the inner regions between them. These observations have been interpreted as evidence for the formation of a current sheet between top of the flare loops and the second coronal source located above the flare loop top \cite{sui2003,sui2005}. The imaging spectroscopy of the two coronal sources was conducted for an M-class flare occurred on April 30, 2002 \cite{liu2008}. In this event, the HXR footpoints were occulted by the limb thus making conditions favorable for the imaging of relatively faint coronal sources. The parameters derived from X-ray spectroscopy of the two sources reveal that the magnetic reconnection site lies between the top of the flare loops and the second coronal source. The formation of double coronal sources, one at each side of the reconnection site, has been viewed in the framework of the stochastic acceleration model \cite{liu2008}. 

\section{Summary}
\label{sec:4}
The multi-wavelength observations have immensely improved our understanding of various physical processes occurring in different atmospheric layers of the Sun during a solar flare. The standard flare model has been successful in broadly recognizing these physical processes as the consequence of large-scale magnetic reconnection in the corona. However, the advancement in the observational capabilities has also led to several new aspects of the flare evolution that deviate from the standard flare model. Although flares have been observed over the past 150 years, but still our understanding about them is incomplete. We have yet to understand several basic elements pertaining to pre-flare magnetic configuration, triggering mechanism, energy release site (e.g., current sheet dimensions), conversion of magnetic energy into heat and kinetic energy, particle acceleration, etc. These outstanding issues pose challenges for future investigations.

\begin{acknowledgement}
This activity has been supported by the European Community Framework Programme 7, `High Energy Solar Physics Data in Europe (HESPE)', grant agreement no.: 263086. We thank Dr. Brajesh Kumar and Dr. Anand Joshi for carefully going through the manuscript. Thanks are also due to an anonymous referee for helpful  suggestions.

\end{acknowledgement}

\bibliographystyle{plain}


\begin{thebibliography}{10}

\bibitem{antiochos1978}
S.~K. {Antiochos} and P.~A. {Sturrock}.
\newblock {Evaporative cooling of flare plasma}.
\newblock {\em \apj}, 220:1137--1143, March 1978.

\bibitem{aschwanden2001}
M.~J. {Aschwanden} and D.~{Alexander}.
\newblock {Flare Plasma Cooling from 30 MK down to 1 MK modeled from Yohkoh,
  GOES, and TRACE observations during the Bastille Day Event (14 July 2000)}.
\newblock {\em \solphys}, 204:91--120, December 2001.

\bibitem{battaglia2009}
M.~{Battaglia}, L.~{Fletcher}, and A.~O. {Benz}.
\newblock {Observations of conduction driven evaporation in the early rise
  phase of solar flares}.
\newblock {\em \aap}, 498:891--900, May 2009.

\bibitem{benz1977}
A.~O. {Benz}.
\newblock {Spectral features in solar hard X-ray and radio events and particle
  acceleration}.
\newblock {\em \apj}, 211:270--280, January 1977.

\bibitem{brosius2003}
J.~W. {Brosius}.
\newblock {Chromospheric Evaporation and Warm Rain during a Solar Flare
  Observed in High Time Resolution with the Coronal Diagnostic Spectrometer
  aboard the Solar and Heliospheric Observatory}.
\newblock {\em \apj}, 586:1417--1429, April 2003.

\bibitem{brown1971}
J.~C. {Brown}.
\newblock {The Deduction of Energy Spectra of Non-Thermal Electrons in Flares
  from the Observed Dynamic Spectra of Hard X-Ray Bursts}.
\newblock {\em \solphys}, 18:489--502, July 1971.

\bibitem{canfield2007}
R.~C. {Canfield}, M.~D. {Kazachenko}, L.~W. {Acton}, D.~H. {Mackay}, J.~{Son},
  and T.~L. {Freeman}.
\newblock {Yohkoh SXT Full-Resolution Observations of Sigmoids: Structure,
  Formation, and Eruption}.
\newblock {\em \apjl}, 671:L81--L84, December 2007.

\bibitem{carmichael1964}
H.~{Carmichael}.
\newblock {A Process for Flares}.
\newblock {\em NASA Special Publication}, 50:451, 1964.

\bibitem{chifor2007}
C.~{Chifor}, D.~{Tripathi}, H.~E. {Mason}, and B.~R. {Dennis}.
\newblock {X-ray precursors to flares and filament eruptions}.
\newblock {\em \aap}, 472:967--979, September 2007.

\bibitem{cliver1986}
E.~W. {Cliver}, B.~R. {Dennis}, A.~L. {Kiplinger}, S.~R. {Kane}, D.~F.
  {Neidig}, N.~R. {Sheeley}, Jr., and M.~J. {Koomen}.
\newblock {Solar gradual hard X-ray bursts and associated phenomena}.
\newblock {\em \apj}, 305:920--935, June 1986.

\bibitem{culhane1970}
J.~L. {Culhane}, J.~F. {Vesecky}, and K.~J.~H. {Phillips}.
\newblock {The Cooling of Flare Produced Plasmas in the Solar Corona}.
\newblock {\em \solphys}, 15:394--413, December 1970.

\bibitem{farnik1996}
F.~{F{\'a}rn{\'{\i}}k}, H.~{Hudson}, and T.~{Watanabe}.
\newblock {Spatial Relations between Preflares and Flares}.
\newblock {\em \solphys}, 165:169--179, April 1996.

\bibitem{farnik1998}
F.~{F{\'a}rn{\'{\i}}k} and S.~K. {Savy}.
\newblock {Soft X-Ray Pre-Flare Emission Studied in Yohkoh-SXT Images}.
\newblock {\em \solphys}, 183:339--357, December 1998.

\bibitem{forbes1996}
T.~G. {Forbes} and L.~W. {Acton}.
\newblock {Reconnection and Field Line Shrinkage in Solar Flares}.
\newblock {\em \apj}, 459:330, March 1996.

\bibitem{glover2000}
A.~{Glover}, N.~D.~R. {Ranns}, L.~K. {Harra}, and J.~L. {Culhane}.
\newblock {The Onset and Association of CMEs with Sigmoidal Active Regions}.
\newblock {\em \grl}, 27:2161, July 2000.

\bibitem{grigis2004}
P.~C. {Grigis} and A.~O. {Benz}.
\newblock {The spectral evolution of impulsive solar X-ray flares}.
\newblock {\em \aap}, 426:1093--1101, November 2004.

\bibitem{hagyard1984}
M.~J. {Hagyard}, D.~{Teuber}, E.~A. {West}, and J.~B. {Smith}.
\newblock {A quantitative study relating observed shear in photospheric
  magnetic fields to repeated flaring}.
\newblock {\em \solphys}, 91:115--126, March 1984.

\bibitem{hirayama1974}
T.~{Hirayama}.
\newblock {Theoretical Model of Flares and Prominences. I: Evaporating Flare
  Model}.
\newblock {\em \solphys}, 34:323--338, February 1974.

\bibitem{hudson1998}
H.~S. {Hudson}, J.~R. {Lemen}, O.~C. {St.~Cyr}, A.~C. {Sterling}, and D.~F.
  {Webb}.
\newblock {X-ray coronal changes during halo CMEs}.
\newblock {\em \grl}, 25:2481--2484, 1998.

\bibitem{ishikawa2011}
S.~{Ishikawa}, S.~{Krucker}, T.~{Takahashi}, and R.~P. {Lin}.
\newblock {On the Relation of Above-the-loop and Footpoint Hard X-Ray Sources
  in Solar Flares}.
\newblock {\em \apj}, 737:48, August 2011.

\bibitem{ji2007}
H.~{Ji}, G.~{Huang}, and H.~{Wang}.
\newblock {The Relaxation of Sheared Magnetic Fields: A Contracting Process}.
\newblock {\em \apj}, 660:893--900, May 2007.

\bibitem{ji2006}
H.~{Ji}, G.~{Huang}, H.~{Wang}, T.~{Zhou}, Y.~{Li}, Y.~{Zhang}, and M.~{Song}.
\newblock {Converging Motion of H{$\alpha$} Conjugate Kernels: The Signature of
  Fast Relaxation of a Sheared Magnetic Field}.
\newblock {\em \apjl}, 636:L173--L174, January 2006.

\bibitem{ji2008}
H.~{Ji}, H.~{Wang}, C.~{Liu}, and B.~R. {Dennis}.
\newblock {A Hard X-Ray Sigmoidal Structure during the Initial Phase of the
  2003 October 29 X10 Flare}.
\newblock {\em \apj}, 680:734--739, June 2008.

\bibitem{joshi2007}
B.~{Joshi}, P.~K. {Manoharan}, A.~M. {Veronig}, P.~{Pant}, and K.~{Pandey}.
\newblock {Multi-Wavelength Signatures of Magnetic Reconnection of a
  Flare-Associated Coronal Mass Ejection}.
\newblock {\em \solphys}, 242:143--158, May 2007.

\bibitem{joshi2005}
B.~{Joshi} and P.~{Pant}.
\newblock {Distribution of H{$\alpha$} flares during solar cycle 23}.
\newblock {\em \aap}, 431:359--363, February 2005.

\bibitem{joshi2006}
B.~{Joshi}, P.~{Pant}, and P.~K. {Manoharan}.
\newblock {North-South Distribution of Solar Flares during Cycle 23}.
\newblock {\em Journal of Astrophysics and Astronomy}, 27:151--157, September
  2006.

\bibitem{joshi2009}
B.~{Joshi}, A.~{Veronig}, K.-S. {Cho}, S.-C. {Bong}, B.~V. {Somov}, Y.-J.
  {Moon}, J.~{Lee}, P.~K. {Manoharan}, and Y.-H. {Kim}.
\newblock {Magnetic Reconnection During the Two-phase Evolution of a Solar
  Eruptive Flare}.
\newblock {\em \apj}, 706:1438--1450, December 2009.

\bibitem{joshi2011}
B.~{Joshi}, A.~M. {Veronig}, J.~{Lee}, S.-C. {Bong}, S.~K. {Tiwari}, and K.-S.
  {Cho}.
\newblock {Pre-flare Activity and Magnetic Reconnection during the Evolutionary
  Stages of Energy Release in a Solar Eruptive Flare}.
\newblock {\em \apj}, 743:195, December 2011.

\bibitem{kim2008}
S.~{Kim}, Y.-J. {Moon}, Y.-H. {Kim}, Y.-D. {Park}, K.-S. {Kim}, G.~S. {Choe},
  and K.-H. {Kim}.
\newblock {Preflare Eruption Triggered by a Tether-cutting Process}.
\newblock {\em \apj}, 683:510--515, August 2008.

\bibitem{kiplinger1995}
A.~L. {Kiplinger}.
\newblock {Comparative Studies of Hard X-Ray Spectral Evolution in Solar Flares
  with High-Energy Proton Events Observed at Earth}.
\newblock {\em \apj}, 453:973, November 1995.

\bibitem{kopp1976}
R.~A. {Kopp} and G.~W. {Pneuman}.
\newblock {Magnetic reconnection in the corona and the loop prominence
  phenomenon}.
\newblock {\em \solphys}, 50:85--98, October 1976.

\bibitem{krucker2008b}
S.~{Krucker}, M.~{Battaglia}, P.~J. {Cargill}, L.~{Fletcher}, H.~S. {Hudson},
  A.~L. {MacKinnon}, S.~{Masuda}, L.~{Sui}, M.~{Tomczak}, A.~L. {Veronig},
  L.~{Vlahos}, and S.~M. {White}.
\newblock {Hard X-ray emission from the solar corona}.
\newblock {\em \aapr}, 16:155--208, October 2008.

\bibitem{krucker2008}
S.~{Krucker}, G.~J. {Hurford}, A.~L. {MacKinnon}, A.~Y. {Shih}, and R.~P.
  {Lin}.
\newblock {Coronal {$\gamma$}-Ray Bremsstrahlung from Solar Flare-accelerated
  Electrons}.
\newblock {\em \apjl}, 678:L63--L66, May 2008.

\bibitem{krucker2008a}
S.~{Krucker} and R.~P. {Lin}.
\newblock {Hard X-Ray Emissions from Partially Occulted Solar Flares}.
\newblock {\em \apj}, 673:1181--1187, February 2008.

\bibitem{kumar2006}
B.~{Kumar} and B.~{Ravindra}.
\newblock {Analysis of Enhanced Velocity Signals Observed during Solar Flares}.
\newblock {\em Journal of Astrophysics and Astronomy}, 27:425--438, December
  2006.

\bibitem{li2005}
Y.~P. {Li} and W.~Q. {Gan}.
\newblock {The Shrinkage of Flare Radio Loops}.
\newblock {\em \apjl}, 629:L137--L139, August 2005.

\bibitem{li2006}
Y.~P. {Li} and W.~Q. {Gan}.
\newblock {The Oscillatory Shrinkage in TRACE 195 {\AA} Loops during a Flare
  Impulsive Phase}.
\newblock {\em \apjl}, 644:L97--L100, June 2006.

\bibitem{lin2002}
R.~P. {Lin}, B.~R. {Dennis}, and G.~J. {Hurford et al.}
\newblock {The Reuven Ramaty High-Energy Solar Spectroscopic Imager (RHESSI)}.
\newblock {\em \solphys}, 210:3--32, November 2002.

\bibitem{lin1976}
R.~P. {Lin} and H.~S. {Hudson}.
\newblock {Non-thermal processes in large solar flares}.
\newblock {\em \solphys}, 50:153--178, October 1976.

\bibitem{lin2003}
R.~P. {Lin}, S.~{Krucker}, G.~J. {Hurford}, D.~M. {Smith}, H.~S. {Hudson},
  G.~D. {Holman}, R.~A. {Schwartz}, B.~R. {Dennis}, G.~H. {Share}, R.~J.
  {Murphy}, A.~G. {Emslie}, C.~{Johns-Krull}, and N.~{Vilmer}.
\newblock {RHESSI Observations of Particle Acceleration and Energy Release in
  an Intense Solar Gamma-Ray Line Flare}.
\newblock {\em \apjl}, 595:L69--L76, October 2003.

\bibitem{liuc2007}
C.~{Liu}, J.~{Lee}, V.~{Yurchyshyn}, N.~{Deng}, K.-S. {Cho}, M.~{Karlick{\'y}},
  and H.~{Wang}.
\newblock {The Eruption from a Sigmoidal Solar Active Region on 2005 May 13}.
\newblock {\em \apj}, 669:1372--1381, November 2007.

\bibitem{liu2010}
R.~{Liu}, C.~{Liu}, S.~{Wang}, N.~{Deng}, and H.~{Wang}.
\newblock {Sigmoid-to-flux-rope Transition Leading to a Loop-like Coronal Mass
  Ejection}.
\newblock {\em \apjl}, 725:L84--L90, December 2010.

\bibitem{liu2009}
R.~{Liu}, H.~{Wang}, and D.~{Alexander}.
\newblock {Implosion in a Coronal Eruption}.
\newblock {\em \apj}, 696:121--135, May 2009.

\bibitem{liu2008}
W.~{Liu}, V.~{Petrosian}, B.~R. {Dennis}, and Y.~W. {Jiang}.
\newblock {Double Coronal Hard and Soft X-Ray Source Observed by RHESSI:
  Evidence for Magnetic Reconnection and Particle Acceleration in Solar
  Flares}.
\newblock {\em \apj}, 676:704--716, March 2008.

\bibitem{manoharan1996}
P.~K. {Manoharan}, L.~{van Driel-Gesztelyi}, M.~{Pick}, and P.~{Demoulin}.
\newblock {Evidence for Large-Scale Solar Magnetic Reconnection from Radio and
  X-Ray Measurements}.
\newblock {\em \apjl}, 468:L73, September 1996.

\bibitem{masuda1994}
S.~{Masuda}, T.~{Kosugi}, H.~{Hara}, S.~{Tsuneta}, and Y.~{Ogawara}.
\newblock {A loop-top hard X-ray source in a compact solar flare as evidence
  for magnetic reconnection}.
\newblock {\em \nat}, 371:495--497, October 1994.

\bibitem{milligan2006b}
R.~O. {Milligan}, P.~T. {Gallagher}, M.~{Mathioudakis}, D.~S. {Bloomfield},
  F.~P. {Keenan}, and R.~A. {Schwartz}.
\newblock {RHESSI and SOHO CDS Observations of Explosive Chromospheric
  Evaporation}.
\newblock {\em \apjl}, 638:L117--L120, February 2006.

\bibitem{milligan2006a}
R.~O. {Milligan}, P.~T. {Gallagher}, M.~{Mathioudakis}, and F.~P. {Keenan}.
\newblock {Observational Evidence of Gentle Chromospheric Evaporation during
  the Impulsive Phase of a Solar Flare}.
\newblock {\em \apjl}, 642:L169--L171, May 2006.

\bibitem{moore1992}
R.~L. {Moore} and G.~{Roumeliotis}.
\newblock {Triggering of Eruptive Flares - Destabilization of the Preflare
  Magnetic Field Configuration}.
\newblock In {Z.~Svestka, B.~V.~Jackson, \& M.~E.~Machado}, editor, {\em IAU
  Colloq. 133: Eruptive Solar Flares}, volume 399 of {\em Lecture Notes in
  Physics, Berlin Springer Verlag}, page~69, 1992.

\bibitem{moore2001}
R.~L. {Moore}, A.~C. {Sterling}, H.~S. {Hudson}, and J.~R. {Lemen}.
\newblock {Onset of the Magnetic Explosion in Solar Flares and Coronal Mass
  Ejections}.
\newblock {\em \apj}, 552:833--848, May 2001.

\bibitem{neupert1968}
W.~M. {Neupert}.
\newblock {Comparison of Solar X-Ray Line Emission with Microwave Emission
  during Flares}.
\newblock {\em \apjl}, 153:L59, July 1968.

\bibitem{nitta2011}
S.~{Nitta}, S.~{Imada}, and T.~T. {Yamamoto}.
\newblock {Clear Detection of Chromospheric Evaporation Upflows with High
  Spatial/Temporal Resolution by Hinode XRT}.
\newblock {\em \solphys}, page 398, November 2011.

\bibitem{pallavicini1977}
R.~{Pallavicini}, S.~{Serio}, and G.~S. {Vaiana}.
\newblock {A survey of soft X-ray limb flare images - The relation between
  their structure in the corona and other physical parameters}.
\newblock {\em \apj}, 216:108--122, August 1977.

\bibitem{pallavicini1990}
R.~{Pallavicini}, G.~{Tagliaferri}, and L.~{Stella}.
\newblock {X-ray emission from solar neighbourhood flare stars - A
  comprehensive survey of EXOSAT results}.
\newblock {\em \aap}, 228:403--425, February 1990.

\bibitem{pandey2012}
J.~C. {Pandey} and K.~P. {Singh}.
\newblock {A study of X-ray flares - II. RS CVn-type binaries}.
\newblock {\em \mnras}, 419:1219--1237, January 2012.

\bibitem{priest2002}
E.~R. {Priest} and T.~G. {Forbes}.
\newblock {The magnetic nature of solar flares}.
\newblock {\em \aapr}, 10:313--377, 2002.

\bibitem{reznikova2010}
V.~E. {Reznikova}, V.~F. {Melnikov}, H.~{Ji}, and K.~{Shibasaki}.
\newblock {Dynamics of the Flaring Loop System of 2005 August 22 Observed in
  Microwaves and Hard X-rays}.
\newblock {\em \apj}, 724:171--181, November 2010.

\bibitem{rust1996}
D.~M. {Rust} and A.~{Kumar}.
\newblock {Evidence for Helically Kinked Magnetic Flux Ropes in Solar
  Eruptions}.
\newblock {\em \apjl}, 464:L199, June 1996.

\bibitem{saldanha2008}
R.~{Saldanha}, S.~{Krucker}, and R.~P. {Lin}.
\newblock {Hard X-ray Spectral Evolution and Production of Solar Energetic
  Particle Events during the January 2005 X-Class Flares}.
\newblock {\em \apj}, 673:1169--1173, February 2008.

\bibitem{schmieder1995}
B.~{Schmieder}, P.~{Heinzel}, J.~E. {Wiik}, J.~{Lemen}, B.~{Anwar}, P.~{Kotrc},
  and E.~{Hiei}.
\newblock {Relation between cool and hot post-flare loops of 26 June 1992
  derived from optical and X-ray (SXT-Yohkoh) observations}.
\newblock {\em \solphys}, 156:337--361, February 1995.

\bibitem{silva2000}
A.~V.~R. {Silva}, H.~{Wang}, and D.~E. {Gary}.
\newblock {Correlation of Microwave and Hard X-Ray Spectral Parameters}.
\newblock {\em \apj}, 545:1116--1123, December 2000.

\bibitem{somov1986}
B.~V. {Somov}.
\newblock {Non-neutral current sheets and solar flare energetics}.
\newblock {\em \aap}, 163:210--218, July 1986.

\bibitem{somov2010}
B.~V. {Somov}.
\newblock {Interpretation of the observed motions of hard X-ray sources in
  solar flares}.
\newblock {\em Astronomy Letters}, 36:514--519, July 2010.

\bibitem{sterling2000}
A.~C. {Sterling}, H.~S. {Hudson}, B.~J. {Thompson}, and D.~M. {Zarro}.
\newblock {Yohkoh SXT and SOHO EIT Observations of Sigmoid-to-Arcade Evolution
  of Structures Associated with Halo Coronal Mass Ejections}.
\newblock {\em \apj}, 532:628--647, March 2000.

\bibitem{sturrock1966}
P.~A. {Sturrock}.
\newblock {Model of the High-Energy Phase of Solar Flares}.
\newblock {\em \nat}, 211:695--697, August 1966.

\bibitem{sui2003}
L.~{Sui} and G.~D. {Holman}.
\newblock {Evidence for the Formation of a Large-Scale Current Sheet in a Solar
  Flare}.
\newblock {\em \apjl}, 596:L251--L254, October 2003.

\bibitem{sui2004}
L.~{Sui}, G.~D. {Holman}, and B.~R. {Dennis}.
\newblock {Evidence for Magnetic Reconnection in Three Homologous Solar Flares
  Observed by RHESSI}.
\newblock {\em \apj}, 612:546--556, September 2004.

\bibitem{sui2005}
L.~{Sui}, G.~D. {Holman}, S.~M. {White}, and J.~{Zhang}.
\newblock {Multiwavelength Analysis of a Solar Flare on 2002 April 15}.
\newblock {\em \apj}, 633:1175--1186, November 2005.

\bibitem{Syrovatskii1972}
S.~I. {Syrovatskii} and O.~P. {Shmeleva}.
\newblock {Heating of Plasma by High-Energy Electrons, and Nonthermal X-Ray
  Emission in Solar Flares}.
\newblock {\em \azh}, 49:334, April 1972.

\bibitem{tappin1991}
S.~J. {Tappin}.
\newblock {Do all solar flares have X-ray precursors?}
\newblock {\em \aaps}, 87:277--302, February 1991.

\bibitem{tsuneta1996}
S.~{Tsuneta}.
\newblock {Structure and Dynamics of Magnetic Reconnection in a Solar Flare}.
\newblock {\em \apj}, 456:840, January 1996.

\bibitem{tsuneta1992}
S.~{Tsuneta}, H.~{Hara}, T.~{Shimizu}, L.~W. {Acton}, K.~T. {Strong}, H.~S.
  {Hudson}, and Y.~{Ogawara}.
\newblock {Observation of a solar flare at the limb with the YOHKOH Soft X-ray
  Telescope}.
\newblock {\em \pasj}, 44:L63--L69, October 1992.

\bibitem{uddin2003}
W.~{Uddin}, B.~{Joshi}, R.~{Chandra}, and A.~{Joshi}.
\newblock {Dynamics of Limb Flare and Associated Primary and Secondary Post
  Flare Loops}.
\newblock {\em Bulletin of the Astronomical Society of India}, 31:303--308,
  March 2003.

\bibitem{venkat2008}
P.~{Venkatakrishnan}, B.~{Kumar}, and W.~{Uddin}.
\newblock {Co-spatial evolution of photospheric Doppler enhancements and
  H{$\alpha$} flare ribbons observed during the solar flare of 2003 October
  28}.
\newblock {\em \mnras}, 387:L69--L73, June 2008.

\bibitem{veronig2004}
A.~M. {Veronig} and J.~C. {Brown}.
\newblock {A Coronal Thick-Target Interpretation of Two Hard X-Ray Loop
  Events}.
\newblock {\em \apjl}, 603:L117--L120, March 2004.

\bibitem{veronig2005}
A.~M. {Veronig}, J.~C. {Brown}, and L.~{Bone}.
\newblock {Evidence for a solar coronal thick-target hard X-ray source observed
  by RHESSI}.
\newblock {\em Advances in Space Research}, 35:1683--1689, 2005.

\bibitem{veronig2006}
A.~M. {Veronig}, M.~{Karlick{\'y}}, B.~{Vr{\v s}nak}, M.~{Temmer},
  J.~{Magdaleni{\'c}}, B.~R. {Dennis}, W.~{Otruba}, and W.~{P{\"o}tzi}.
\newblock {X-ray sources and magnetic reconnection in the X3.9 flare of 2003
  November 3}.
\newblock {\em \aap}, 446:675--690, February 2006.

\bibitem{veronig2010}
A.~M. {Veronig}, J.~{Ryb{\'a}k}, P.~{G{\"o}m{\"o}ry}, S.~{Berkebile-Stoiser},
  M.~{Temmer}, W.~{Otruba}, B.~{Vr{\v s}nak}, W.~{P{\"o}tzi}, and
  D.~{Baumgartner}.
\newblock {Multiwavelength Imaging and Spectroscopy of Chromospheric
  Evaporation in an M-class Solar Flare}.
\newblock {\em \apj}, 719:655--670, August 2010.

\bibitem{vrsnak2006}
B.~{Vr{\v s}nak}, M.~{Temmer}, A.~{Veronig}, M.~{Karlick{\'y}}, and J.~{Lin}.
\newblock {Shrinking and Cooling of Flare Loops in a Two-Ribbon Flare}.
\newblock {\em \solphys}, 234:273--299, April 2006.

\bibitem{warren2001}
H.~P. {Warren} and A.~D. {Warshall}.
\newblock {Ultraviolet Flare Ribbon Brightenings and the Onset of Hard X-Ray
  Emission}.
\newblock {\em \apjl}, 560:L87--L90, October 2001.

\bibitem{zarro1988}
D.~M. {Zarro} and J.~R. {Lemen}.
\newblock {Conduction-driven chromospheric evaporation in a solar flare}.
\newblock {\em \apj}, 329:456--463, June 1988.

\bibitem{zhang2001}
J.~{Zhang}, K.~P. {Dere}, R.~A. {Howard}, M.~R. {Kundu}, and S.~M. {White}.
\newblock {On the Temporal Relationship between Coronal Mass Ejections and
  Flares}.
\newblock {\em \apj}, 559:452--462, September 2001.

\end{thebibliography}
\end{document}